# Two-dimensional connective nanostructures of electrodeposited Zn on Au(111) induced by spinodal decomposition

J. Dogel, R. Tsekov and W. Freyland

Institute of Physical Chemistry, University of Karlsruhe (TH), D-76128 Karlsruhe, Germany

Phase-formation of surface alloying by spinodal decomposition has been studied for the first time at an electrified interface. For this aim Zn was electrodeposited on Au(111) from the ionic liquid $AlCl_3$-MBIC (58:42) containing 1 mM Zn(II) at different potentials in the underpotential range corresponding to submonolayer up to monolayer coverage. Structure evolution was observed by *in situ* electrochemical scanning tunneling microscopy (STM) at different times after starting the deposition via potential jumps and at temperatures of 298 K and 323 K. Spinodal or labyrinth two-dimensional structures predominate at middle coverage, both in deposition and dissolution experiments. They are characterized by a length scale of typically 5 nm which has been determined from the power spectral density of the STM images. Structure formation and surface alloying is governed by slow kinetics with a rate constant $k$ with activation energy of 120 meV and pre-exponential factor of 0.17 Hz. The evolution of the structural features is described by a continuum model and is found to be in good agreement with the STM observations. From the experimental and model calculation results we conclude that the two-dimensional phase-formation in the Zn on Au(111) system is dominated by surface alloying. The phase separation of a Zn-rich and a Zn-Au alloy phase is governed by 2D spinodal decomposition.

In considering first order phase transitions and subsequent phase formation the effect of spinodal reactions on the phase morphology commonly is not taken into account. This is the case both with bulk systems and with interfaces or electrified interfaces[1]. Phase formation in these systems usually is governed by nucleation and growth phenomena, where the thermodynamic aspects have a long tradition going back to the work of Gibbs. Mass transport in this case is described by normal diffusion, i.e. down the concentration gradient. However, in systems with small diffusivity like some solid alloys, glasses or viscous fluids, the whole two-phase region defined by the spinodal can be accessible. This was first shown by Cahn and Hilliard who developed the theory of spinodal decomposition and verified it in a number of simple solid and glassy systems[2-4]. Phase separation within the spinodal follows a thermodynamically modified diffusion equation whereby concentration fluctuations lead to a flux in the direction up the concentration gradient. The corresponding diffusional clustering mechanism is quite different from that of conventional nucleation and growth. Small clusters of approximately nanometer size can arrange nearly periodically in space leading to highly connective structures. This is the essential diagnostic that phase separation occurs by a spinodal mechanism[3].

An example of a 3D system exhibiting spinodal decomposition is the Zn-Al alloy which has been discussed in detail by Cahn[4]. The evolution of a connective or labyrinth structure has been followed by small angle X-ray scattering from which a characteristic length scale of ~5 nm has been determined as expected for the spinodal diffusion mechanism. More recently, a few observations of labyrinth structures (sometimes also called wormlike structures) have been reported at surfaces and interfaces[5-8]. In homoepitaxial growth of Cu/Cu(100) and Ag/Ag(100) adlayers 2D wormlike nanoclusters have been characterized by scanning tunneling microscopy (STM)[5]. The adlayer coverage typically was ~0.6 monolayer (ML). The reshaping of these clusters by pinch-off phenomena is described within a continuum model by peripheral diffusion. First evidence of nanocrystal self-assembly by fluid spinodal decomposition at a fluid/wall interface has been published by Ge and Brus[6]. Drying thin wet films of an organic solvent containing monodisperse CdSe nanocrystals on smooth graphite (HOPG) substrate, the authors find labyrinth patterns with varying coverage. This is explained by an increased van der Waals interaction between the nanocrystals in air as compared within the solution. Two investigations of spinodal decomposition at electrified interfaces have been reported so far. Schuster et al have performed a fascinating experiment of electrochemical dissolution of Au adatoms at a Au(111) surface by applying microsecond voltage pulses. The surface structure was monitored by *in situ* STM and at Au coverage between 0.4 and 0.9 ML labyrinth patterns with typical length scales of ~4 nm were observed[8]. In a second experiment, Erlebacher et al studied the morphology of Ag-Au alloy surfaces after selective dissolution of Ag by electrochemical etching. Again, patterns typical of spinodal decomposition with length scales of 5 – 10 nm are found. Modeling the evolution of the nanoporosity of the surface on etching by a continuum model including the Cahn-Hilliard diffusion yields good agreement with experiment[7].

First indications of labyrinth structures in electrodeposition of Ge and Zn on Au(111) from ionic liquids have been obtained recently in our group by *in situ* electrochemical STM investigations[9,10]. From visual inspection of the STM images of 2D Ge deposits on Au(111) wormlike structures with a length scale of ~5 nm become clear. In this context it is interesting to note that Ge on a close-packed metal surface tends to phase separation and induces surface alloying at low coverage[11]. In Zn/Au(111) deposition surface alloying is manifested after *in situ* dissolution by holes in the Au(111) terraces of 1 to 2 monolayer depth[10].

In this study we have continued our STM investigations of Zn electrodeposition and dissolution in the underpotential deposition (UPD) range. Under these conditions up to 3 monolayers of Zn can be grown on Au(111) from an ionic liquid[10]. We have focused on the time evolution of the relevant labyrinth structures after jumping the potentials into the deposition and dissolution ranges, respectively. Structural information has been derived from the power spectra of the respective STM images. They are compared with the predictions of a fluid continuum model which are in good quantitative agreement with the experimental results, both in respect with the time dependence and the length scales of the spinodal structures.

The experimental results reported here have been obtained with the same *in situ* STM technique and under similar electrochemical conditions as described previously so that we may refer to this publication for further details[10]. For the electrodeposition of Zn on Au(111) an ionic liquid electrolyte, aluminium chloride-1-methyl-3-butylimidazolium chloride (AlCl$_3$-BMIC, 58:42) has been employed with 1 mM Zn(II) dissolved anodically. This electrolyte has a large enough electrochemical window to avoid hydrogen evolution during Zn electrodeposition so that this influence on the morphology of the deposit is excluded. *In situ* STM measurements were performed with a home-built microscope which allows measurements under pure Ar atmosphere and at variable temperature; its construction is similar to the design described previously[12]. The microscope was driven by a Digital Instrument Nanoscope E controller in combination with a Molecular Imaging picostat. Electrochemically etched tungsten tips of 0.25 mm diameter insulated by an epoxide painting were used for room temperature measurements. At elevated temperatures glass insulated Pt/Ir (90:10) tips were necessary which were prepared by melting suitable glass capillaries with the Pt/Ir wire[13]. STM measurements have been combined with *in situ* and *ex situ* experiments of cyclic voltammetry and chronoamperometry. However, in this study we focus on the STM imaging of the evolution of 2D labyrinth nanostructures of Zn on Au(111) during electrodeposition and dissolution. For this aim we have followed the STM images after potential jumps in the UPD range, typically from 750 mV to the respective deposition potential and back on dissolution.

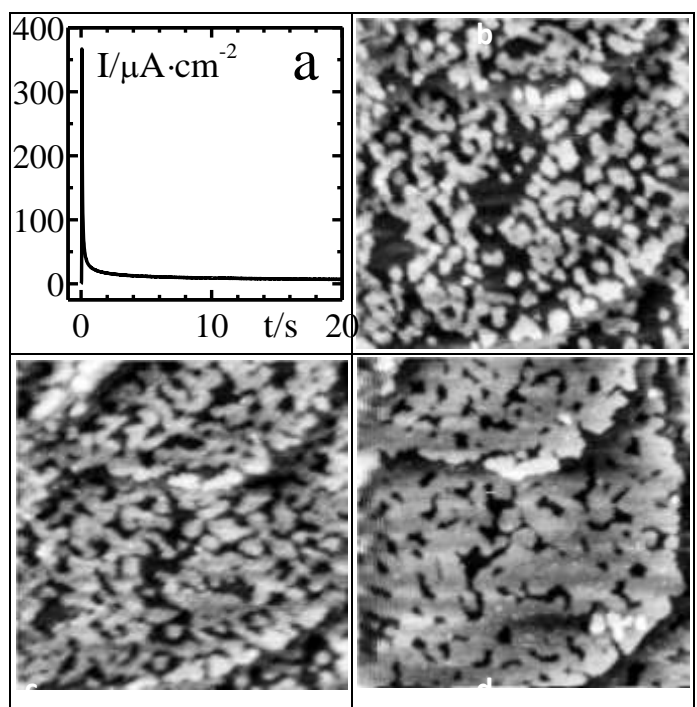

**Fig. 1.** STM observation of evolution of labyrinth structures during the first monolayer deposition of Zn on Au(111) from $AlCl_3$-BMIC+1 mM $Zn^{2+}$ in the UPD range after a potential jump from 750 mV to 200 mV; a) current transient; b–d) STM images (50 x 50 nm²) taken in the time interval $8 \leq t \leq 81$ min after the potential jump, scan rate 3.9 Hz, $I_{tum}$ = 5 nA, $E_{tip}$ = 100 mV

**Fig. 2.** Evolution of labyrinth structures during dissolution of Zn deposited on Au(111); STM images (129 x 129 nm²) have been taken at different times after a potential jump from -50 mV to 750 mV; a) 3 min, b) 8 min and c) 90 min; electrodeposition and STM parameters similar to those of Fig. 1

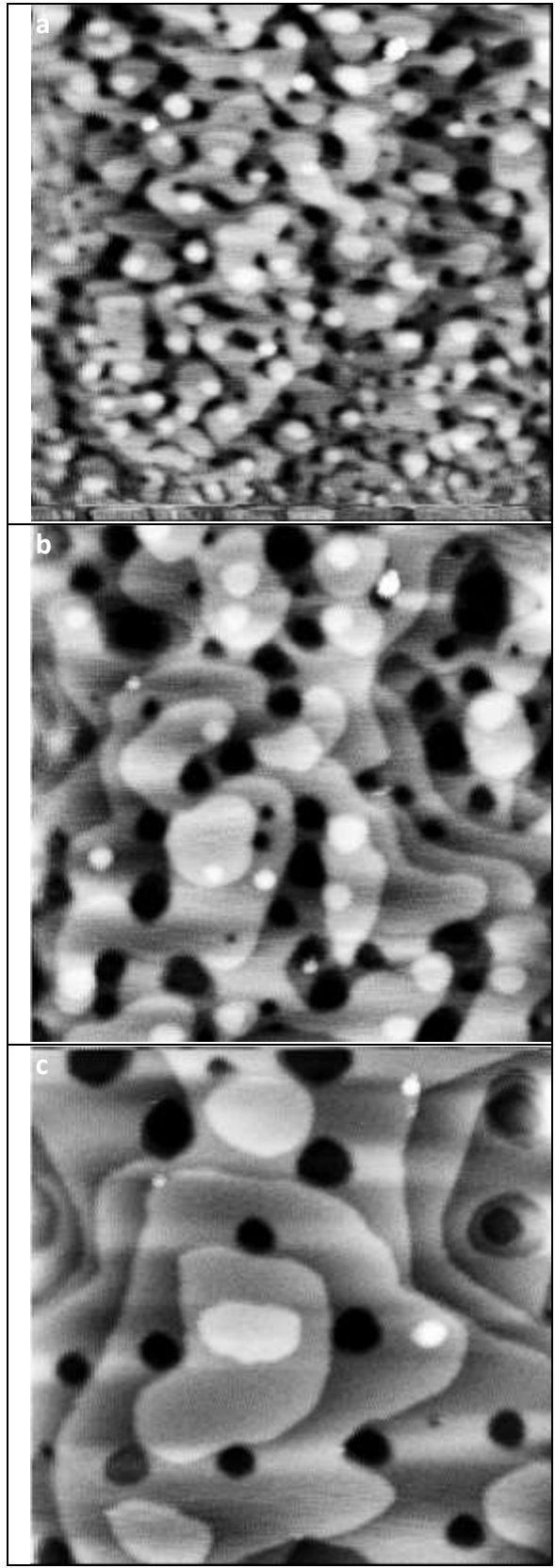

In previous experiments we have shown that Zn underpotential deposition in the ionic liquid occurs via a layer-by-layer mechanism where up to 3 monolayers form at UPD conditions[10]. This corresponds to three reduction waves in the cyclic voltammogram, with a first one weakly indicated between 300 and 200 mV followed by two reduction peaks at 100 and 10 mV, respectively. A few selected STM images representing the characteristic changes of the surface morphology during electrodeposition of the first Zn monolayer are depicted in Fig. 2 in Ref. [10]. As a reference the image of the pure Au(111) terraces taken at 300 mV is shown in part a). Island formation is clearly visible in image b) at 250 mV; images taken at this potential at different time indicate an instantaneous nucleation and growth mechanism. Connective or labyrinth structures are observed at potentials around 200 mV, see part c). Reducing the potential further to 160 mV the formation of the first monolayer is completed with exception of a few vacancy islands. In a very similar fashion the morphology changes during the formation of a second monolayer (150 – 60 mV) and a third one (60 – 0 mV). This layer-by-layer growth mechanism continues in the OPD range[10]. From STM measurements the thickness of the first layer is determined to be $2.2 \pm 0.2$ Å. With respect to the following, two observations are of particular interest. First, with respect to pure Au the effective tunneling barrier of Zn in the first monolayer is reduced by ~0.5 eV, whereas that of the second and third monolayer is clearly less reduced, by ~0.25 eV [13]. Secondly, in none of the completed Zn monolayers on Au(111) a superstructure leading to a Moirée pattern has been observed. This is typically seen in systems not exhibiting surface alloying. An example is Ni on Au(111), which has a lattice misfit similar to that of Zn on Au(111) [14,15].

Fig. 1 shows the evolution of a monolayer thick labyrinth structure of Zn on Au(111) imaged at different times after a potential jump from 750 mV to 200 mV. The corresponding current transient, Fig. 1a, exhibits a nearly exponential decay at short times with a relaxation time of about $0.2$ s; integration of the $I$ vs. $t$ curve yields a coverage corresponding to $0.1-0.2$ ML. At longer times a small current in the µA range seems to flow which may contribute a small change in the coverage, see below. In the time interval studied the STM images indicate a transformation of the labyrinth structures towards higher connectivity. After about 80 minutes wormlike vacancy patterns with typical pinch-off contours are left. A very similar behavior has been observed during electrodeposition of the second and also the third monolayer.

A selection of STM images representing the structural changes at the electrode/electrolyte interface on electrochemical dissolution of Zn/Au(111) is given in Fig. 2. In this case more than 3 monolayers of Zn had been deposited at -50 mV (OPD range) and the potential was jumped to an anodic value of 750 mV. Wormlike structures are visible in the beginning of the dissolution process which after more than one hour transform into the typical Au-substrate leaving holes of up to 3 monolayer depth. In cases where only UPD films are dissolved the depth of the holes is reduced to 1 to 2 monolayers. They heal after several hours whereby the respective Au islands – see white spots in the STM images – disappear. The holes appearing on dissolution we interpret as a fingerprint of surface alloying, which is clearly visible also after dissolution of the first deposited monolayer.

The variation of the average surface coverage $\Theta$ of the labyrinth structures with time has been evaluated from the STM images with the aid of the DI software 4.23 and a Fortran routine. These results are plotted in Fig. 3 as a function of time for different stepping potentials relating to the first monolayer formation (a) and the second monolayer (b). In both cases – for the step to 160 mV and to 80 mV, respectively – a complete coverage of $\Theta \sim 1$ is reached after about 10 min. The time dependence of these curves follows an exponential law, i.e.

$$\Theta = 1 - \exp(-kt),\tag{1}$$

which is shown for the example of the curve for 200 mV for two different temperatures in Fig. 4. Such a rate low is expected for a reversible reaction between lattice incorporated and desorbed Zn atoms whereby the rate constant k describes the slow structural transformation or surface alloying process[16]. The respective rate constants are $k = (1.6 \pm 0.2) \times 10^{-3}$ s$^{-1}$ at 298 K, $k = (2.3 \pm 0.2) \times 10^{-3}$ s$^{-1}$ at 323 K. Assuming Arrhenius behavior for $k$ vs. $T$, the activation energy of 120 meV is estimated from these values.

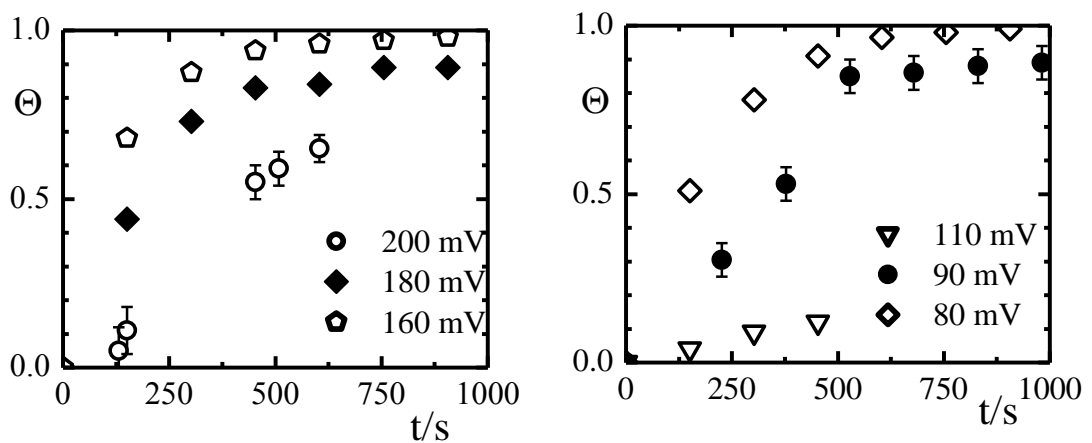

**Fig. 3.** Time dependence of the average coverage $\Theta$ of the labyrinth structure at different stepping potentials; a) first Zn monolayer formation, b) second Zn monolayer formation

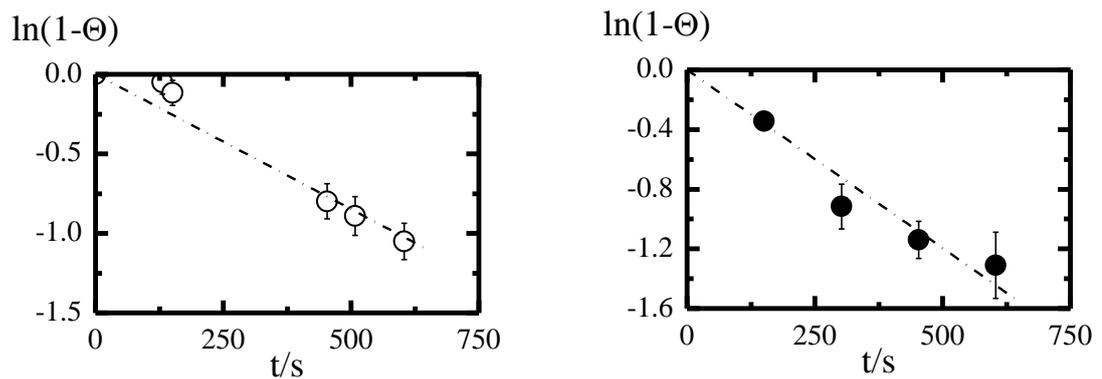

**Fig. 4.** Logarithmic plot of average coverage $\Theta$ versus t of first monolayer electrodeposition at 200 mV for a) 298 K and b) 323 K

In order to evaluate the characteristic length scales of the labyrinth structures the STM images have been analyzed by the DI Nanoscope software DI 4.23 which gives the so called power spectral density. In principle, it presents the dominant Fourier components of the surface structure and is proportional to the structure factor $S_q$. A typical result of this evaluation is shown in

Fig. 5 at two different temperatures. It develops with time and at long times decays again, see Fig. 5a. The maximum of the dependence of $S_q$ on $\lambda$ defines the relevant wavelength $\lambda_m$ of the surface structure. In both cases a dominant structure with a length scale of ~5 nm is apparent. In determining the power spectral densities from the STM images care was taken to scan the structures only over flat sections of the images, avoiding step edges, which could contribute an artificial component to the power spectral density.

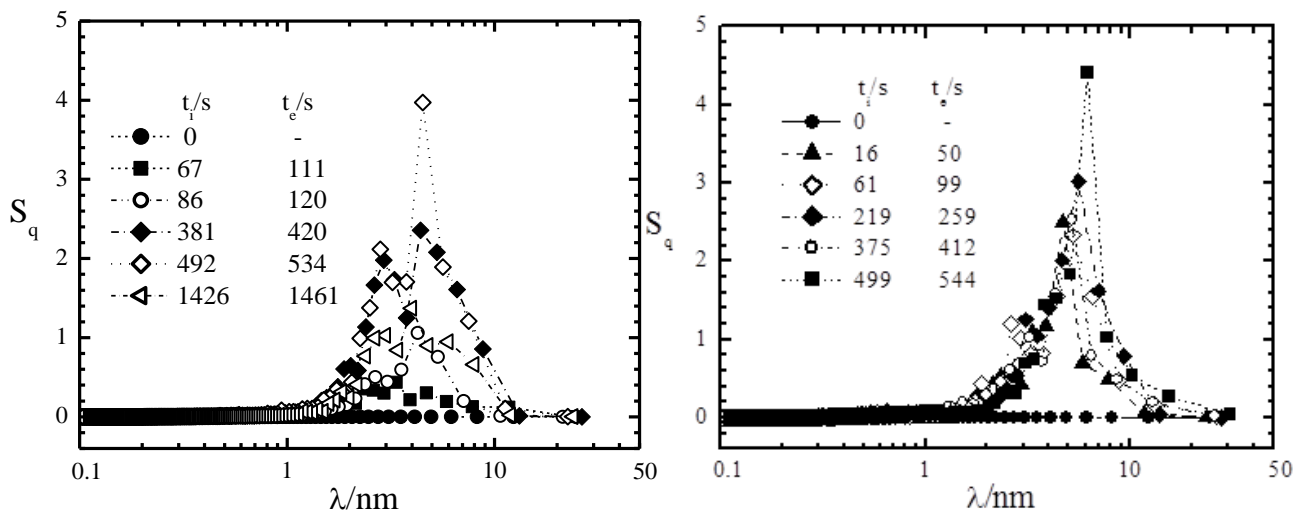

**Fig. 5.** Structure factor $S_q$ versus the wave length $\lambda$ of the labyrinth structure of the first Zn monolayer deposition at 200 mV and a) 298 K and b) 323 K; plotted are the $S_q$ data at different times, where the intervals $t_i \leq t \leq t_e$ denote the times of scanning the STM images after the deposition pulse

The experimental results show that the two-dimensional phase formation of electrodeposited Zn on Au(111) is characterized by a labyrinthine surface structure with typical length scales of ~5 nm. This structure develops by a relatively slow surface reaction with rate constants of the order of $10^{-3}$ s$^{-1}$. The labyrinth structure, its length scale and the slow dynamics at the interface are strong indications of a spinodal demixing reaction occurring at the electrified interface. To our knowledge there is no theoretical description for this case. Therefore we have developed a simple hydrodynamic model to get further insight into the microscopic structure and dynamics of the two-dimensional phase formation driven by spinodal decomposition at the electrified interface. It is described in the following together with the main model predictions.

Since the Zn atoms in the dilute surface phase are very mobile it is reasonable to believe that the surface alloying takes place only at surface parts covered by condensed surface phase, where the adatoms are practically laterally fixed. Hence, the surface alloying affects the properties of the condensed surface phase, which dynamics is, however, hidden for our STM measurements. Regardless of the actual location and state of the zinc atoms on the gold surface, their thermodynamic properties can be effectively described by the Gibbs adsorption $\Gamma$ defined on the equimolecular gold surface, where the adsorption of gold is zero. In the range of the applied electrical potentials the adsorption of other species from the AlCl$_3$-MBIC liquid is negligible[17].

Hence, if the order parameter $\phi \equiv \sigma^2 \Gamma$ is introduced with $\sigma = 2.7$ Å being the diameter of a Zn or Au atom, the 2D hydrodynamics of the idealized one-component Zn fluid adsorbed on the gold surface is governed via the following mass and momentum conservation balances

$$\partial_t \phi + \nabla_s \cdot (\phi \mathbf{v}_s) = \sigma^2 j_n \tag{2}$$

$$\nabla_s \cdot (m\phi \xi_s \nabla_s \mathbf{v}_s) + \nabla_s (m\phi \zeta_s \nabla_s \cdot \mathbf{v}_s) - \phi \beta_s \mathbf{v}_s = \phi \nabla_s \mu_s \tag{3}$$

$$\mu_s = \mu_s^0 + k_B T[\ln(\phi) - \ln(1-\phi) - w\phi - \kappa \sigma^2 \nabla_s^2 \phi] + \delta \mu_f \tag{4}$$

Here $\nabla_s$ is the 2D nabla operator, $\mathbf{v}_s$ is the surface hydrodynamic velocity, $j_n$ is the normal flux of Zn atoms from the bulk, $m$ is the atomic mass of zinc, $\beta_s$ is the friction coefficient of the Zn atoms and $\mu_s^0$ is a standard surface chemical potential. Since in general the 2D Zn fluid is not a Newtonian one, the interfacial kinematic shear $\xi_s$ and dilatational $\zeta_s$ viscosities could depend on the adsorption and shear rate. Due to high surface friction the inertial terms and the friction of Zn atoms with the bulk liquid are neglected in Eq. (3). It is important to note that there are two different contributions to the energy dissipation of the Zn atoms: internal fluid friction described by the first two terms in Eq. (3) and the friction with the solid surface described by the third term of Eq. (3).

The chemical potential $\mu_s$ of the Zn atoms in the adsorbed layer corresponds in Eq. (4) to the Bragg-Williams-Frumkin lattice gas model extended by a gradient term and the thermal fluctuation $\delta \mu_f$. The parameters $w$ of attraction between the Zn atoms and $\kappa$ being proportional to the line tension between the surface phases can be calculated via the relations[2]

$$w = (2\pi/\sigma^2) \int_\sigma^\infty [\exp(-U/k_B T) - 1] r \, dr \tag{5}$$

$$\kappa = (\pi/2\sigma^4) \int_\sigma^\infty [\exp(-U/k_B T) - 1] r^3 \, dr \tag{6}$$

if the interaction potential $U$ between the Zn atoms in the adsorbed layer is known. It is convenient to describe the Zn atoms as hard spheres, which attract each other via a force mediated by the gold electron gas[18]

$$U = -\varepsilon \cos(2k_F r)(\sigma/r)^5 \tag{7}$$

Here $k_F = 1.2$ Å$^{-1}$ is the Fermi wave vector of the bulk gold and $\varepsilon$ is a constant of coupling between the gold and zinc. Note that the Friedel oscillations on the surface decay much stronger with the distance between the adatoms as compared to the bulk. The expression (7) takes into account only the effect of the bulk gold electrons since the extremely long-range interaction, induced by the surface gold electrons and being inverse proportional to $r^2$, has negligible contribution[18]. In general, the interaction between the gold and zinc atoms is effectively included not only in $U$ but also in all other specific parameters of the equations above.

The experimental results show that the flux $j_n$ is determined by the adsorption/desorption mechanism. The driving force of the adsorption is the difference of the electrochemical potentials of $Zn^{2+}$ in the bulk and of Zn atoms adsorbed on the surface. Since the diffusion in the bulk is very fast, the relaxation of the electrochemical potential of $Zn^{2+}$ ions in the bulk is very quick. Hence, from the viewpoint of the surface processes the bulk electrochemical potential remains constant. Therefore, only the changes of the surface chemical potential $\mu_s$ determine the adsorption/desorption dynamics. In this case the normal flux of Zn atoms can be expressed as follows

$$\sigma^2 j_n = \alpha(\mu_s)(1-\phi) \tag{8}$$

where the function $\alpha$ decreases with increase of $\mu_s$ to $\alpha = 0$ at equilibrium. The kinetic factor $(1-\phi)$ takes into account the fact that the adsorption takes place only on a bare surface. The latter is confirmed by the experimentally observed layer-by-layer growth of the Zn adsorption.

It is difficult to solve in general the non-linear system of Eqs. (2-4). A possibility to simplify the problem is to linearize them. If $\bar{\phi}$ denotes the average relative density, one can consider the local deviation $\delta\phi = \phi - \bar{\phi}$ of the order parameter, $\delta\mu_f$ and the surface hydrodynamic velocity $\mathbf{v}_s$ as small parameters. Thus, after linearization the equations above acquire the forms

$$\partial_t \bar{\phi} = \bar{\alpha}(1-\bar{\phi}) \tag{9}$$
$$\partial_t \delta\phi + \bar{\phi}\nabla_s \cdot \mathbf{v}_s = \bar{\alpha}'(1-\bar{\phi})\delta\mu_s \tag{10}$$
$$m\bar{\xi}_s \nabla_s^2 \mathbf{v}_s + m\bar{\zeta}_s \nabla_s(\nabla_s \cdot \mathbf{v}_s) - \bar{\beta}_s \mathbf{v}_s = \nabla_s \delta\mu_s \tag{11}$$
$$\bar{\mu}_s = \mu_s^0 + k_B T[\ln(\bar{\phi}) - \ln(1-\bar{\phi}) - w\bar{\phi}] \tag{12}$$
$$\delta\mu_s = k_B T[\delta\phi/\bar{\phi}(1-\bar{\phi}) - w\delta\phi - \kappa\sigma^2 \nabla_s^2 \delta\phi] + \delta\mu_f \tag{13}$$

where $\bar{\alpha} \equiv \alpha(\bar{\mu}_s)$ and $\bar{\alpha}' \equiv (\partial_{\mu_s}\alpha)_{\bar{\mu}_s}$ is a negative quantity.

Let us first consider the evolution of the average adsorption of Zn atoms on the gold surface. If the gold surface is almost empty, the chemical potential $\bar{\mu}_s$ tends to zero. Hence, the adsorption of a dilute layer of Zn atoms is very fast. Later on, if the Zn fluid undergoes a 2D phase separation, according to Eq. (12) the chemical potential $\bar{\mu}_s$ is more or less constant during the transition. If in addition $\alpha$ is a weak function of $\mu_s$ in the two-phase region one can consider $\bar{\alpha}$ as a constant independent of $\bar{\phi}$. Thus, solving Eq. (9) the time-dependence of the average order parameter during the phase transition yields the form

$$\bar{\phi} = 1 - (1-\phi_0)\exp(-\bar{\alpha}t) \tag{14}$$

where $\phi_0$ is the low relative density loaded initially during the fast adsorption before the surface phase transition to begin. The initial value $\phi_0$ can be approximated by the lower value of the order parameter corresponding to the two-phase equilibrium, i.e. $\bar{\mu}_s(\phi_0) = \bar{\mu}_s(1-\phi_0)$. Since the average order parameter is related to the coverage via $\bar{\phi} = (1-\Theta)\phi_0 + \Theta(1-\phi_0)$, the exponential dependence in Eq. (14) corresponds well to the experimental observations on $\Theta$ summarized in Eq. (1) and hence $\bar{\alpha} = k$. The relatively low value of $\bar{\alpha}$ estimated from the experiments can be attributed to surface alloying, which exhibits a rate constant of the same order.

It is convenient to analyze the density perturbations via 2D special Fourier components with wave vector $q$. Expressing the surface hydrodynamic velocity from Eq. (11) as a function of the fluctuations of the surface chemical potential and introducing the result in Eq. (10) yields the following dynamic equation for the evolution of the Fourier component of the density perturbations

$$\partial_t \delta\phi_q = -[\bar{\phi}q^2/(\bar{\beta}_s + m\bar{\xi}_s q^2 + m\bar{\zeta}_s q^2) - \bar{\alpha}'(1-\bar{\phi})]\delta\mu_s^q \equiv -M_q \delta\mu_s^q / 2k_B T \tag{15}$$

where for convenience the mobility $M_q$ is introduced. The first term of $M_q$ describes the surface diffusion of Zn atoms, while the second term accounts for the adsorption/desorption mechanism of propagation of the surface density fluctuations. As was mentioned, however, the experiment shows that $\bar{\alpha}$ is nearly constant in the two-phase region. This implies that $\bar{\alpha}'$ is very small and, for this reason, the contribution of the adsorption/desorption mechanism to the mobility is negligible. This also can be attributed to the surface alloying, which because of the strong Zn-Au interaction (the enthalpy of mixing of the bulk liquid is -177 meV)[19] suppresses dramatically the desorption rate.

In general, the averaged coefficients $\bar{\beta}_s$, $\bar{\xi}_s$ and $\bar{\zeta}_s$ take into account both the transport in the fluid surface phase and the dynamics of incorporation of Zn atoms in the 2D solid phase. Since the experiments show that the dynamics of the two-phase decomposition is much slower than the regular diffusion on metal surfaces[20], the rate determining step should be the peripheral diffusion of Zn atoms along the boundary separating the fluid and solid 2D-phases. Hence, the effective peripheral diffusion coefficient can be expressed as follows

$$2k_B T /(\bar{\beta}_s + m\bar{\xi}_s q^2 + m\bar{\zeta}_s q^2) = (1-\bar{\phi})\nu_0 \sigma^2 \exp(-w - \kappa\sigma^2 q^2) \tag{16}$$

where $\nu_0$ is the attempt frequency. The activation energy of the peripheral diffusion is modeled in Eq. (16) as the energy of consecutive incorporation/expulsion of Zn atoms on the line of two-phase contact properly modified for the curvature effect. The factor $(1-\bar{\phi})$ in Eq. (16) accounts for the fact that if the whole surface is covered by the solid phase the diffusivity is zero. One can expand the right hand side of Eq. (16) in series if $\kappa\sigma^2 q^2 < 1$ and thus identify the specific properties of the 2D Zn fluid

$$\bar{\beta}_s = 2k_B T \exp(w)/v_0\sigma^2(1-\bar{\phi}) \tag{17}$$

$$\bar{\xi}_s + \bar{\zeta}_s = 2k_B T \kappa \exp(w)/m v_0(1-\bar{\phi}) \tag{18}$$

Note that according to Eq. (17), $\bar{\beta}_s$ represents the friction acting on a mobile Zn atom moving along the immobilized atoms in the solid phase. Since the potential $U$ is strongly mediated by the gold surface, $w$ accounts also for the Zn-Au interaction. The surface kinematic viscosities from Eq. (18) are proportional to the line tension, thus representing the specific friction of the motion of the contact line dividing the two surface phases. Substituting Eqs. (17) and (18) in the mobility, the latter changes to

$$M_q = \bar{\phi}(1-\bar{\phi})v_0\sigma^2 q^2 \exp(-w)/(1+\kappa\sigma^2 q^2) \tag{19}$$

Substituting Eq. (13) in Eq. (15) yields a stochastic differential equation describing the dynamics of the small adsorption perturbations

$$2\partial_t \delta\phi_q + M_q \Omega_q \delta\phi_q = -M_q \delta\mu_f^q / k_B T \tag{20}$$

where $\Omega_q \equiv 1/\bar{\phi}(1-\bar{\phi}) - w + \kappa\sigma^2 q^2$ is the susceptibility of the adsorption to energy fluctuations. The last term in Eq. (20) represents the random Langevin force causing the permanent Brownian motion of the interfacial structure. While the mobility $M_q$ is always positive, the susceptibility $\Omega_q$ can change its sign. For large $q$, $\Omega_q$ is positive, which according to Eq. (20) results in stable density fluctuations. If $w > 1/\bar{\phi}(1-\bar{\phi})$ $\Omega_q$ becomes negative by decreasing of $q$. In this case spinodal decomposition of the adsorbed Zn fluid will take place. The critical wave number $q_c$ calculated from the relation $\Omega_q = 0$ equals to

$$q_c = \sqrt{[w - 1/\bar{\phi}(1-\bar{\phi})]/\kappa\sigma^2} \tag{21}$$

In the range of unstable Fourier components there is strong competition between the mobility and susceptibility. While the mobility $M_q$ increases the susceptibility $\Omega_q$ decreases by increase of $q$. As a result the product $M_q \Omega_q$ has a minimum corresponding to the fastest unstable Fourier component. The corresponding wave number $q_m$ can be calculated from $\partial_q(M_q \Omega_q) = 0$

$$q_m = \sqrt{[\sqrt{1 + w - 1/\bar{\phi}(1-\bar{\phi})} - 1]/\kappa\sigma^2} \tag{22}$$

In the case of weak interaction, i.e. $0 < w - 1/\bar{\phi}(1-\bar{\phi}) \ll 1$, $q_m$ reduces to the classical result $q_c/\sqrt{2}$ of the Cahn-Hilliard theory[2-4], where the surface viscosity effect is neglected. Since $q_c/\sqrt{2}$ is the upper limit for the wave vector $q_m$, the surface viscosity effect results in a shift of the typical wave vectors of the surface structures to lower values as compared to the Cahn-Hilliard theory, which is experimentally observed[8].

Since Eq. (20) is a stochastic one any solution of it is a random distribution of the adsorption perturbations. For comparison with experiments it is necessary to obtain the evolution of an average quantity. In the present case a convenient statistical moment is the following dimensionless spectral density or structure factor

$$S_q(t) = <\delta\phi_q \delta\phi_q^*> A/\sigma^2 \qquad (23)$$

where $A$ is the area of the sample. Multiplying Eq. (20) by $\delta\theta_q^*$, adding the complex conjugated result and taking the average of the products yields the following equation for the dynamics of the spectral density

$$\partial_t S_q + M_q \Omega_q S_q = M_q \qquad (24)$$

In the derivation of Eq. (24) the well-known equipartition theorem, stating that any degree of freedom possesses fluctuation energy $k_B T$, is employed, i.e. $-<\delta\mu_f^q \delta\phi_q^*> A/\sigma^2 = k_B T$. It is easy to solve Eq. (24) under the initial condition $S_q(t=0) = 0$ and the solution is

$$S_q = \int_0^t M_q(t_1) \exp(-\int_{t_1}^t M_q(t_2)\Omega_q(t_2) dt_2) dt_1 \qquad (25)$$

Note that $M_q$ and $\Omega_q$ depend on time via the time-dependence of $\bar{\phi}$ given by Eq. (14).

To calculate the evolution of the structure factor $S_q$ from Eq. (25) several parameters are required. As was mentioned, the relaxation constant $\bar{\alpha}$ can be taken from the experimental measurements of the evolution of the average coverage $\Theta$ of the surface. Hence, the only variable parameters left are the bonding energy $\varepsilon$ and the attempt frequency $\nu_0$. In Fig. 6 the evolution of the spectral density at $T = 298$ K is presented for $\bar{\alpha} = 1.6$ mHz, $\varepsilon = 82$ meV and $\nu_0 = 19$ Hz. As seen the structure factor exhibits a pronounced maximum corresponding to spinodal decomposition on the surface. However, since the average coverage $\bar{\phi}$ permanently changes in time, far from the maximum the susceptibility $\Omega_q$ is positive and the structure factor drops down. The structure factor in Fig. 6 corresponds well to the experimental observations in Fig. 5 if one accepts that the STM signal corresponds to the local adsorption. The fitted value of $\varepsilon$ is close to the kink formation energy 74 meV of gold[8] but it is four times larger than the typical Zn-

Zn bond energy[21]. This strongly mediated by gold interaction can lead to surface alloying, which can also explain the very low value of the fitted attempt frequency. In Fig. 6 the evolution of the spectral density at $T = 323$ K is presented for $\bar{\alpha} = 2.3$ mHz. The parameters $\varepsilon = 82$ meV and $\nu_0 = 19$ Hz are the same since they do not depend on temperature. As seen the increase of the temperature accelerates the structure formation. The acceleration effect is due to faster $\bar{\phi}$ evolution. The spinodal decomposition, however, is much slower, since the temperature is closer to the critical one. The latter equals to $T_c = 344$ K according to the relation $w(T_c) = 4$ and $\varepsilon = 82$ meV. Indeed, the experimental observation shows a very slow evolution of the resonant peaks, which is typical for the so-called critical slowdown. The wave length $\lambda_m$ corresponding to the maximum of the structure factor defines the length scale of the surface structure. According to Fig. 7 $\lambda_m$ increases in time as a result of the increase of the average density $\bar{\phi}$. Note that only the middle part of the evolution of $\lambda_m$ is affected by spinodal decomposition, while the later stages correspond to usual growth of the order parameter at low supersaturation to $\lambda_m \to \infty$ at $\bar{\phi} \to 1$. The juxtaposition between theory and experiment is good and the discrepancy at 323 K could be attributed to non-linear effects, which increase simultaneously with temperature.

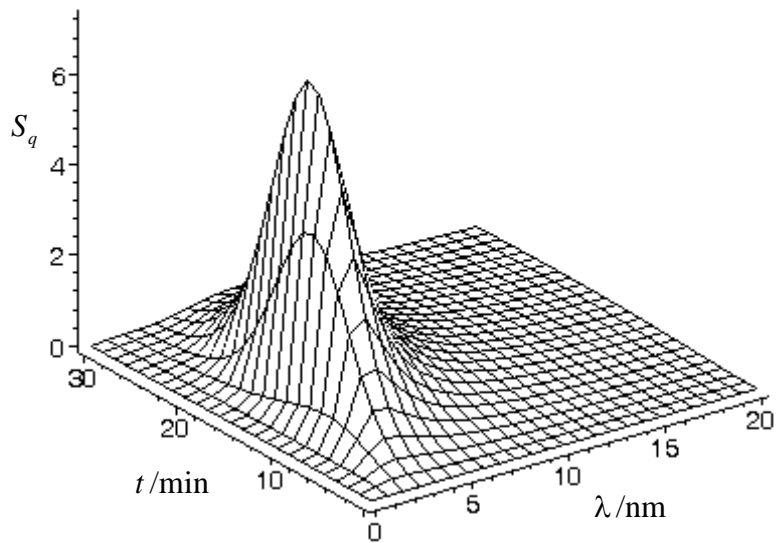

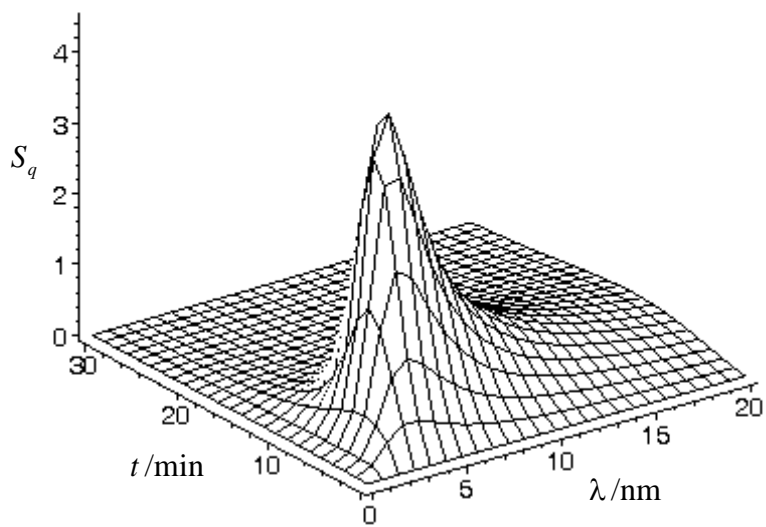

**Fig. 6.** Calculated structure factor $S_q$ versus the wave length $\lambda$ of the labyrinth structure and time $t$ of the first Zn monolayer at 200 mV and 298 K (top) or 323 K (bottom)

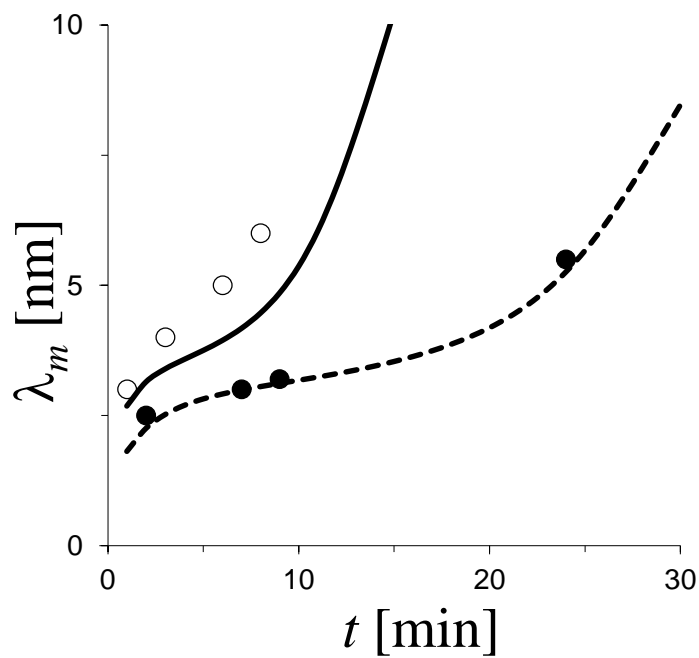

**Fig. 7.** Theoretical evolution of the typical wavelength $\lambda_m$ of the labyrinth structure on time at 200 mV and 298 K (dash line) or 323 K (solid line).
The experimental points (● 298 K and ○ 323 K) are estimated from Fig. 6

An essential prediction of the theory of spinodal decomposition is the occurrence of two-phase connective structures[3]. They develop in the spinodal region by fluctuations of density or composition with specific wavelengths. The dominant wavelengths are solely determined by the thermodynamic properties of the systems, the curvature of the free energy and the strength and length scale of the interatomic interaction[4]. In one-component bulk fluid systems with typical diffusivities in the range $10^{-5}$ cm²/s, the spinodal decomposition is fast with a time scale of microseconds. On the other hand, for solid alloys where the critical demixing temperature is well below the melting point, the time constants of spinodal reactions can reach values of the order of minutes or hours. A particular feature of the two-phase spinodal structures is their striking connectivity leading to labyrinth patterns. These are distinct from the morphologies expected for nucleation and growth.

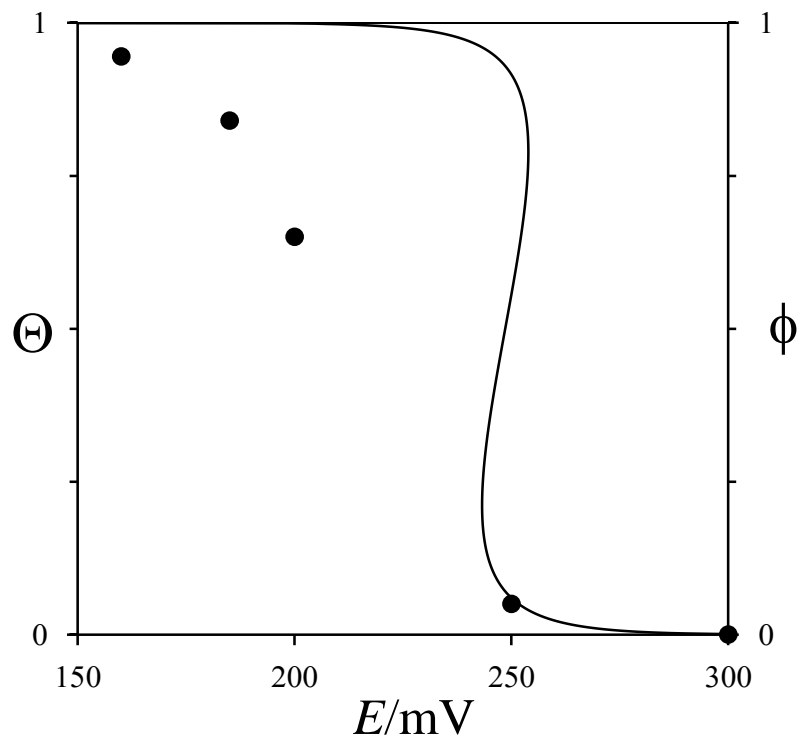

**Fig. 8.** Dependence of the order parameter $\phi$ from the model calculations and the average coverage $\Theta(t = 600\,\text{s})$ of the labyrinth structure (the experimental points) on the applied electric potential $E$

The STM observations of the electrodeposition of Zn on Au(111) from the ionic liquid strongly indicate that first order two-dimensional phase transitions occur in the underpotential (UPD) range. This is demonstrated in Fig. 8 for the example of the first monolayer formation. Plotted is the order parameter $\phi$ as a function of the deposition potential $E$, calculated by the balance of the average surface chemical potential from Eq. (12) and the electrochemical potential of $Zn^{2+}$ ions in the bulk, as well as the $\Theta$ values at long polarization times of 600 s. Near a potential of 250 mV, $\phi$ vs. $E$ exhibits a shape as expected for a first order phase transition. The latter is evident from the low non-zero value of $\Theta$. The STM results give evidence, that the two-dimensional phase separation is governed by a spinodal mechanism below 200 mV. The STM images exhibit typical interconnective structures, the dominant length scale is in the range of ~5 nm. Hence, the strong increase of the coverage $\Theta$ at 200 mV in Fig. 7 can be associated to a transition from the slow nucleation and growth mechanism to a faster one. According to Cahn's diagnostic a continuous change of the coverage is considered as a proof of spinodal decomposition[3]. The question arises: what is the origin of the 2D spinodal structures. Two possibilities are of interest. In the first case one may consider a first order transition from a fluid phase of Zn adatoms to a condensed 2D film where the spinodal separation is driven by density fluctuations. For this mechanism to hold the slow dynamics corresponding to the measured rate constant $k \sim 10^{-3}$ s$^{-1}$ has to be explained. The model calculations focus on this problem. The second possibility is that surface alloying dominates and the transition between a 2D Zn-rich phase and a 2D Zn-Au alloy phase is driven by concentration fluctuations. In this case the rate determining step is the incorporation of Zn into Au which is a slow diffusional process.

Disregarding surface alloying at the moment, the kinetics of metal UPD on a substrate is influenced by a number of factors, including bulk diffusion of metal ions, the charge transfer reaction, surface diffusion of metal adatoms, surface inhomogeneities like kink sites and step edges, and finally, nucleation and growth of the 2D phases[22]. An important parameter is the strength of adsorbed metal-substrate interaction in comparison with the metal-metal interaction. Both non-monotonous and monotonous current transients are found if 2D nucleation and growth steps contribute in the 2D metal phase formation process[22]. In our measurements the current transients in the UPD range exhibit a monotonous behavior, see the example in Fig. 1. In our model calculations relatively low peripheral and lattice incorporation diffusion coefficients of ~$10^{-17}$ cm$^2$/s have to be assumed in order to get agreement with the experimental structure factor. For comparison, surface diffusion coefficients of metal atoms on fcc metal surfaces range from e.g. ~$10^{-6}$ cm$^2$/s for Ag/Ag(111) to $10^{-19}$ cm$^2$/s for Pt/Rh(001) at room temperature[20]. At electrified interfaces, the surface diffusion of metal atoms can be faster than in vacuum due to lowering of the activation energies[23]. However, we could not find a value for the diffusion coefficient of Zn on Au(111) in an electrochemical environment.

Electrochemical investigations of the kinetics of surface alloying have been performed for the system Pb on Ag(111) at variable temperatures [16, 24-26]. The rate determining step is the incorporation of an adsorbed metal adatom into the metal substrate surface layer. Considering a reversible incorporation and a slow desorption reaction a relation similar to Eq. (1) is derived for the surface alloy coverage as a function of time[16]. For Pb on Ag(111) surface alloying a rate constant $k \sim 10^{-3}$ s$^{-1}$ has been determined at room temperature[16], which is directly comparable with the value of Zn on Au(111) of this study. Assuming that exchange of adatoms occurs only across the first or second layer of the Au(111) substrate this results in a diffusion coefficient of ~$10^{-18}$ cm$^2$/s. From the temperature dependence of $k$ (see Fig. 4) we estimate an activation energy of the diffusion coefficient of 120 meV. It is in reasonable agreement with the value of 300 meV evaluated for the Pb on Ag(111) system[25]. So the kinetics of the surface coverage of Zn on Au(111) can be explained by a surface alloying mechanism. Independent evidence of this mechanism is seen in the one to two monolayer deep holes in the Au(111) substrate which occur on dissolution, see Fig. 2. In summary, we conclude that the two-dimensional phase transition in the system Zn on Au(111) is governed by surface alloying. The phase separation of a Zn-rich and a Zn-Au-alloy phase is controlled by spinodal decomposition which leads to the characteristic labyrinth structures.